\documentclass[cameraready]{Interspeech}
\usepackage{multirow}  
\usepackage{tabularx}

\title{UBG-Net: An Uncertainty-aware Bayesian Gating Network for Robust Audio-Visual Speech Recognition}

\author[affiliation={1}]{Jinjie}{Fu}
\author[affiliation={1}, correspondingauthor]{Hang}{Chen}
\author[affiliation={1}]{Wu}{Guo}
\author[affiliation={1}]{Zhijun}{Zhang}
\author[affiliation={1}]{Kuiliang}{Li}
\author[affiliation={1}]{Peng}{Gao}

\address{
    $^{\text{1}}$NERC-SLIP, University of Science and Technology of China, Hefei, China
}
\email{jjfu@mail.ustc.edu.cn, hangchen@ustc.edu.cn}

\begin{document}

\maketitle

\begin{abstract}
    Audio-Visual speech recognition systems often degrade in real-world scenarios due to signal corruption and distribution shifts. To address this, we propose a unified uncertainty-modeling framework, namely the uncertainty-aware Bayesian gating network (UBG-Net). UBG-Net features a Modality Uncertainty-aware Bayesian Fusion (MUBF) mechanism that injects signal-level aleatoric uncertainty into a Bayesian network to model epistemic uncertainty, thereby ensuring robust fusion of pre-trained backbone features. For inference, we introduce Distribution Uncertainty-aware Hierarchical Voting (DUHV) to select transcripts from Monte Carlo samples, prioritizing frequency and using inference scores in case of a tie. Experiments on the AVCocktail and LRS2 datasets demonstrate the overall superiority of UBG-Net compared to SOTA baselines. Ablation studies confirm that MUBF and DUHV effectively filter noise, enhancing fusion and decoding robustness.
\end{abstract}

\section{Introduction}

Audio-Visual Speech Recognition (AVSR) leverages visual cues, such as lip movements, to compensate for audio degradation\cite{michelsanti2021overview, mcgurk1976hearing}. This mechanism is particularly crucial in noisy environments where acoustic signals are corrupted, yet visual information remains reliable. However, most existing AVSR models rely on fixed audio–visual fusion paradigms~\cite{ivanko2023review, ma2023auto, rouditchenko24_interspeech, li2023robust, zhao2024amg}, thereby limiting their generalization to dynamic real-world scenarios, such as cross-modal heterogeneity (e.g., fluctuating audio-visual reliability)~\cite{nguyen25b_interspeech}.

To mitigate such degradation, the core challenge lies in effectively evaluating and utilizing the reliability of each modality. Traditional approaches primarily rely on attention mechanisms or gating modules to dynamically weight audio and visual features~\cite{rouditchenko24_interspeech, li2023robust, hu2023cross}. While effective under matched conditions, these deterministic methods yield only point estimates and inherently fail to capture input reliability. Recent works have thus begun to explore probabilistic modeling to capture two distinct types of uncertainty: aleatoric uncertainty (inherent data noise) and epistemic uncertainty (model knowledge limitation)~\cite{kendall2017uncertainties, gawlikowski2023survey}. However, existing attempts in multimodal tasks typically treat these uncertainties in isolation~\cite{zhu2025proxy, gao2024embracing, subedar2019uncertainty, huang2025latent, tellamekala2023cold}, overlooking their potential correlation. For instance, studies like~\cite{zhu2025proxy, gao2024embracing} focus primarily on modeling aleatoric uncertainty via variance learning, whereas the approach in~\cite{subedar2019uncertainty} models epistemic uncertainty through Bayesian Neural Networks (BNNs)~\cite{blundell2015weight}. Unfortunately, such isolated practices implicitly assume these uncertainties are independent. This assumption fails to capture the intrinsic dependency between data quality and model confidence. From our perspective, severe signal corruption should be explicitly accounted for in estimating epistemic uncertainty, ensuring that data noise is not misconstrued as semantic ambiguity.

\begin{figure*}[t]
  \centering
  \includegraphics[width=1.0\linewidth]{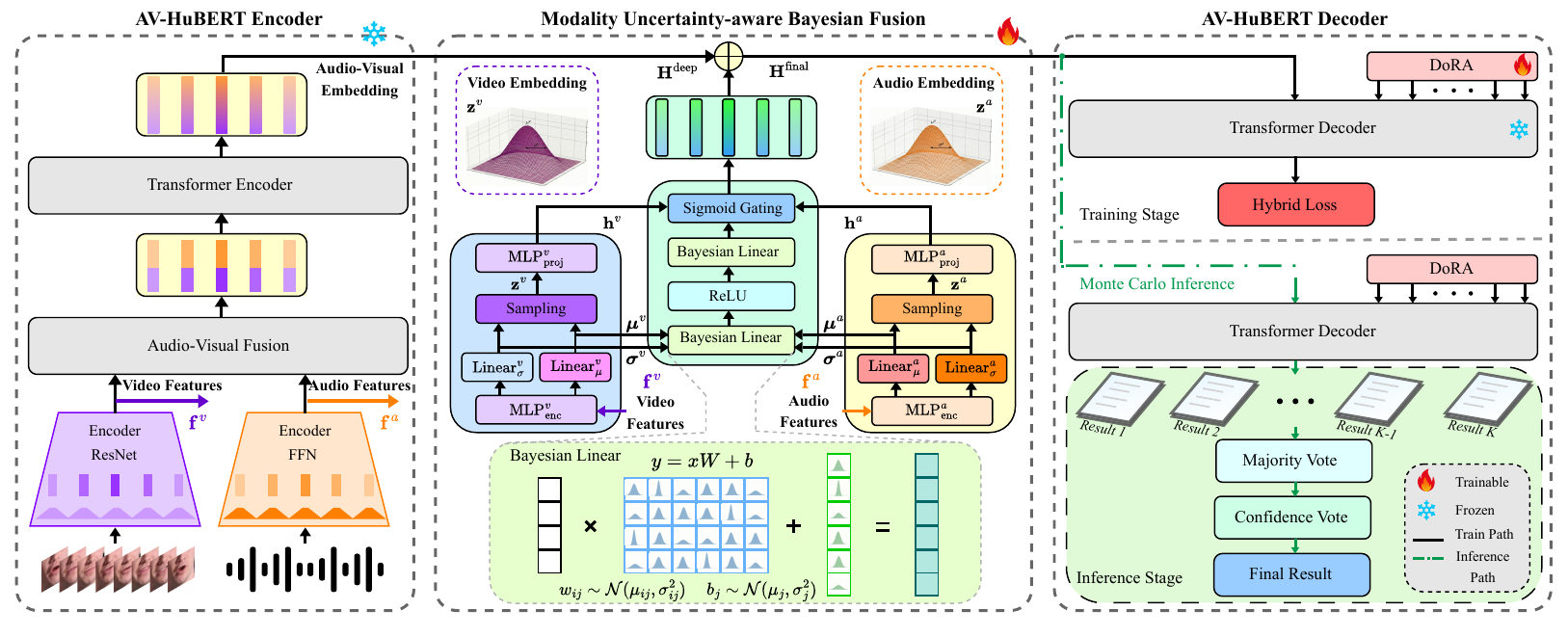}
  \caption{Framework Overview. (1) AV-HuBERT Encoder: A pre-trained backbone employed for deep multimodal representation extraction. (2) Modality Uncertainty-aware Bayesian Fusion: A Bayesian gating network that modulates multimodal streams using injected aleatoric uncertainty parameters. (3) AV-HuBERT Decoder: The decoding module, where the top and bottom parts denote the training and inference stages, respectively.}
  \label{fig:arch}
\end{figure*}

In this paper, we propose a novel uncertainty-aware Bayesian gating network (UBG-Net) to enhance AVSR performance under challenging conditions. Instead of treating data noise and model ambiguity as separate entities, we design a modality uncertainty-aware Bayesian fusion~(MUBF) mechanism in which signal-level aleatoric uncertainty serves as a contextual cue for model-level epistemic uncertainty. Specifically, we explicitly model the input distribution to capture aleatoric uncertainty and inject this information into a Bayesian gating network, which dynamically modulates the uncertainty-aware features with the pre-trained backbone representations. Furthermore, we introduce a distribution uncertainty-aware hierarchical voting~(DUHV) strategy to address the sequence alignment challenge inherent in the Monte Carlo (MC) sampling~\cite{malinin2021uncertainty} employed during inference to approximate the posterior predictive distribution~\cite{kendall2017uncertainties}. Specifically, we first apply majority voting to hypotheses generated via MC sampling and retain the most frequent candidates. When multiple candidates tie for the highest frequency, the final prediction is selected based on the highest inference score. In summary, the main contributions of  UBG-Net are:
\begin{itemize}
    \item proposing the MUBF mechanism that utilizes signal-level aleatoric uncertainty as context to explicitly guide the estimation of model-level epistemic uncertainty, offering a more elaborate control over multimodal fusion.
    \item designing the DUHV strategy to resolve alignment ambiguities in Monte Carlo sampling, which selects the optimal hypothesis by first applying majority voting and then leveraging inference scores in case of a tie.
   \item performing extensive experiments across the challenging AVCocktail dataset~\cite{nguyen25b_interspeech} and the noise-augmented LRS2 benchmarks~\cite{son2017lip}, achieving new state-of-the-art~(SOTA) performances and demonstrating excellent generalization ability \footnote{We did not include LRS3 in our evaluation as the dataset was unavailable for access at the time of this research.}.
\end{itemize}

\section{Proposed UBG-Net}
The proposed UBG-Net for AVSR, as depicted in Fig.~\ref{fig:arch}, includes three parts: the AV-HuBERT encoder for multimodal deep representation extracting, the modality uncertainty-aware Bayesian fusion block and the AV-HuBERT decoder. Our innovations focus on the latter two parts, and we will provide detailed descriptions in the following sections.

\subsection{Modality uncertainty-aware Bayesian fusion}
\label{sec:uum}

We propose a unified framework to jointly model aleatoric and epistemic uncertainties, as depicted in the middle of Fig.~\ref{fig:arch}.

\subsubsection{Aleatoric uncertainty encoders}

To capture aleatoric uncertainty inherent in the data, we map features into a stochastic latent space through modality-dependent encoders. For the input frame-level feature $\mathbf{f}^m_t$, the mean $\boldsymbol{\mu}^m_t$ and standard deviation $\boldsymbol{\sigma}^m_t$ of the distribution can be calculated as
\begin{equation}
  \left[ \boldsymbol{\mu}^m_t, \log (\boldsymbol{\sigma}^m_t)^2 \right] = \mathrm{Linear}^m_{\{\mu, \sigma\}}\left( \mathrm{MLP}^m_{\mathrm{enc}}(\mathbf{f}^m_t) \right)
  \label{eq:prob_encoder}
\end{equation}
where $\mathrm{MLP}^m_{\mathrm{enc}}$ and $\mathrm{Linear}^m_{\{\mu, \sigma\}}$ stand for multi-layer perception module and the following parallel linear heads. In this and the following sections, the superscripts or subscripts $m \in \{v, a\}$ denote the modality (visual or audio) and $t \in \{1, \dots, T\}$ index the time frames within an input sequence of length $T$. $(\boldsymbol{\sigma}^m_t)^2$ quantifies the observed data heteroscedasticity, and we predict $\log (\boldsymbol{\sigma}^m_t)^2$ because it is more numerically stable than regressing the variance $(\boldsymbol{\sigma}^m_t)^2$ in neural network training~\cite{kendall2017uncertainties}. Instead of point estimates, we generate stochastic embeddings $\mathbf{z}^m_t$ via the reparameterization trick~\cite{kingma2014auto}:

\begin{equation}
  \mathbf{z}^m_t = \boldsymbol{\mu}^m_t + \boldsymbol{\epsilon} \odot \boldsymbol{\sigma}^m_t, \quad \boldsymbol{\epsilon} \sim \mathcal{N}(\mathbf{0}, \mathbf{I})
  \label{eq:reparam}
\end{equation}
Finally, the refined features $\mathbf{h}^m_t = \mathrm{MLP}^m_{\mathrm{proj}}(\mathbf{z}^m_t)$ are used for the subsequent information fusion. A KL divergence loss $\mathcal{L}_{\mathrm{KL\text{-}Feat}}$ constrains this latent space towards a standard Gaussian prior~\cite{alemi2017deep}.

\subsubsection{Epistemic uncertainty gating}

Unlike previous approaches that model aleatoric and epistemic uncertainty independently, we design a Bayesian Gating Network (BGN) in which aleatoric uncertainty serves as a contextual cue for modeling epistemic uncertainty, guiding the gating process and effectively bridging the two types of uncertainty. As shown in Fig.~\ref{fig:arch}, the BGN consists of two Bayesian linear layers~\cite{blundell2015weight}, a ReLU activation, and a sigmoid gating function. For each Bayesian linear layer, we place a variational distribution $q_\theta(\mathbf{W})$ over the weights and define the prior distribution as a standard Gaussian $p(\mathbf{W}) = \mathcal{N}(\mathbf{W}; \mathbf{0}, \mathbf{I})$~\cite{graves2011practical}. The inputs to the bottom Bayesian linear layer are not raw features but the distribution parameters. We construct a perceptual context vector $\mathbf{c}_t$ by concatenating the mean and variance from both modalities:
\begin{equation}
    \mathbf{c}_t = \operatorname{Concat}\left(\left[\boldsymbol{\mu}_t^v, (\boldsymbol{\sigma}_t^v)^2, \boldsymbol{\mu}_t^a, (\boldsymbol{\sigma}_t^a)^2\right]\right)
    \label{eq:context}
\end{equation}

The context $\mathbf{c}_t$ feeds into the next Bayesian linear layers to produce a stochastic gating vector $\mathbf{g}_t$, which dynamically modulates the multimodal features for information fusion.
\begin{align}
    \mathbf{g}_t &= \sigma\Big( \operatorname{BNN}(\mathbf{c}_t; \mathbf{W}) \Big) \label{eq:gating} \\
    \mathbf{H}_t^{\mathrm{final}} &= \mathbf{H}_t^{\mathrm{deep}} + \mathbf{g}_t \odot \operatorname{Concat}(\mathbf{h}_t^v, \mathbf{h}_t^a) \label{eq:fusion}
\end{align}
where $\mathbf{H}_t^{\mathrm{deep}}$ is the deep representation from a pre-trained model, $\operatorname{BNN}(\cdot)$ denotes the composite transformation of two Bayesian linears with an intermediate ReLU.

\begin{table}[t]
  \centering
  \small  
  \caption{WER (\%) comparison on the LRS2 dataset. Note that 'Int.' denotes the number of interfering speakers.}
  \label{tab:lrs2_results_fixed}
  
  \setlength{\tabcolsep}{6pt}
  
  \begin{tabular}{l c ccccc c} 
    \toprule
    \multirow{2}{*}{\textbf{Model}} & \multirow{2}{*}{\textbf{Int.}} & \multicolumn{5}{c}{\textbf{SNR (dB)}} & \multirow{2}{*}{\textbf{Avg}} \\
    \cmidrule(lr){3-7}
     & & \textbf{-5} & \textbf{0} & \textbf{5} & \textbf{10} & $\infty$ & \\
    \midrule
    
                 & 0 &  &  &  &  & 2.1 & \\
    Baseline     & 1 & 6.4 & 3.5 & 3.4 & 2.8 &  & 4.1 \\
                 & 2 & 9.0 & 4.4 & 3.2 & 2.8 &  & \\
    \midrule
    
                 & 0 &  &  &  &  & 4.0 & \\
    Whisper      & 1 & 96.6 & 30.4 & 14.1 & 6.0 &  & 33.8 \\
                 & 2 & 104.0 & 30.4 & 13.1 & 5.4 &  & \\
    \midrule
    
\multirow{3}{*}{\shortstack{Qwen3\\-Omni}} & 0 &  &  &  &  & 4.2 & \\
                                           & 1 & 96.4 & 30.0 & 13.5 & 5.6 &  & 32.8 \\
                                           & 2 & 100.0 & 30.1 & 10.8 & 4.8 &  & \\
    \midrule
    
                 & 0 &  &  &  &  & 2.2 & \\
    UBG-Net & 1 & 5.2 & 3.1 & 3.1 & 2.6 &  & 3.9 \\
                 & 2 & 8.3 & 4.1 & 3.2 & 3.0 &  & \\
    \bottomrule
  \end{tabular}%
  
\end{table}

To learn the weight posterior, we adopt the Variational Inference (VI) method, which optimizes the negative Evidence Lower Bound (ELBO)~\cite{blundell2015weight}. Within this objective, the Bayesian layers contribute the complexity cost, formulated as the KL divergence between the variational posterior and the prior:
\begin{equation}
  \mathcal{L}_{\mathrm{KL\text{-}Weight}} = \mathrm{KL}(q_\theta(\mathbf{W}) || p(\mathbf{W}))
  \label{eq:kl_weight}
\end{equation}
Mathematically, the summation of the recognition loss (acting as the negative log-likelihood) and this KL term constitutes the standard negative ELBO for the model parameters.

To ensure stable convergence, we apply the local reparameterization trick~\cite{kingma2015variational}. Instead of sampling weights directly, we sample the pre-activation outputs, thereby decoupling the sample variance from the minibatch size. 

Finally, the total loss combines the negative ELBO (comprising the recognition loss and weight priors) with explicit regularization on the latent feature space:
\begin{equation}
  \mathcal{L}_{\mathrm{total}} = \mathcal{L}_{\mathrm{CTC/Attn}} + \beta_1 \mathcal{L}_{\mathrm{KL\text{-}Weight}} + \beta_2 \mathcal{L}_{\mathrm{KL\text{-}Feat}}
  \label{eq:total_objective}
\end{equation}
where $\mathcal{L}_{\mathrm{CTC/Attn}}$ is the weighted sum of CTC and attention losses. The coefficients $\beta_1$ and $\beta_2$ balance the model complexity cost and the feature-level constraints, respectively.

\subsection{Distribution uncertainty-aware hierarchical voting}
\label{sec:MC inference}

During inference, we propose a distribution-uncertainty-aware hierarchical voting (DUHV) strategy to effectively leverage the diversity of predictions produced by the Bayesian gating network. Specifically, we employ Monte Carlo (MC) sampling to approximate the posterior predictive distribution\cite{blundell2015weight}:
\begin{equation}
  p(\mathbf{y}|\mathbf{x}) \approx \frac{1}{K} \sum_{k=1}^{K} p(\mathbf{y}|\mathbf{x}, \mathbf{W}_k), \quad \mathbf{W}_k \sim q_\theta(\mathbf{W})
\end{equation}

Since sequence lengths vary across $K$ hypotheses by the MC method in the inference stage, standard averaging is inapplicable~\cite{malinin2021uncertainty}. Then, let $\mathcal{H} = \{(\mathbf{y}_k, s_k)\}_{k=1}^K$ be the set of hypotheses, where $s_k$ denotes the inference score, defined as the sequence log-probability obtained via beam search decoding. For each unique hypothesis $\mathbf{y}$, we calculate its frequency $N(\mathbf{y})$ and peak confidence $S_{\mathrm{peak}}(\mathbf{y}) = \max_{k: \mathbf{y}_k=\mathbf{y}} s_k$. The optimal sequence $\mathbf{y}^*$ is determined by a hierarchical selection rule. First, we apply majority voting and obtain the set of candidates with the maximum frequency.
\begin{equation}
  \mathcal{Y}_{\mathrm{cand}} = \{ \mathbf{y} \mid N(\mathbf{y}) = \max_{\mathbf{y}' \in \mathcal{H}} N(\mathbf{y}') \}
\end{equation}  

If the candidate set has only one sequence, this sequence is the final transcript. In the event of a tie, we select the one with the highest inference score from the candidates by the first step
\begin{equation}
  \mathbf{y}^* = \underset{\mathbf{y} \in \mathcal{Y}_{\mathrm{cand}}}{\arg\max} \ S_{\mathrm{peak}}(\mathbf{y})
\end{equation}

\begin{table}[htbp]
  \centering
  \small
  \caption{Ablation study of different components and  different inference strategies in UBG-Net on the AVCocktail dataset, where "w/o" denotes "without".}
  \label{tab:wer_and_ablation_study_1}
  
  \begin{tabular}{l c c c}
    \toprule
    \multirow{3}{*}{\textbf{Model}} & 
    \multicolumn{3}{c}{\textbf{Segmentation}} \\
    
    \cmidrule(lr){2-4}
    
     & \textbf{Active Speaker} & \textbf{Fixed} & \multirow{2}{*}{\textbf{Gold}} \\
     & \textbf{Detection} & \textbf{chunk (10s)} & \\
    \midrule

    Baseline & 22.6 & 39.2 & 18.2 \\
    \midrule
    Whisper & 67.3 & 136.0 & 52.3 \\

    Qwen3-Omni & 70.5 & 143.3 & 54.9 \\
    \midrule

    UBG-Net & 21.9 & 36.1 & 17.4 \\
    
    \quad w/o Epistemic & 22.3 & 37.8 & 17.9 \\
    
    \quad w/o Aleatoric & 22.1 & 37.6 & 17.7 \\

    \cmidrule(l{1.5em}){1-4}
    
    \quad w/o Majority & 22.4 & 37.5 & 17.7 \\
    
    \quad w/o Confidence & 22.2 & 37.8 & 17.8 \\
    
    \bottomrule
  \end{tabular}
\end{table}

\section{Experimental Setup}

\subsection{Datasets and metrics}

For training, we use a combination of datasets including LRS2, Vox2, and AVYT~\cite{nguyen25b_interspeech}.

For evaluation, we use two benchmarks:

\begin{itemize}
    \item \textbf{Simulated LRS2~\cite{son2017lip}:} Following the protocol of Nguyen et al.~\cite{nguyen25b_interspeech}, we introduce up to two background interfering speakers and apply different signal-to-noise ratios (SNRs) of $\{-5, 0, 5, 10\}$ dB. This setup measures robustness against controlled synthetic interruptions.
    \item \textbf{Real-world AVCocktail~\cite{nguyen25b_interspeech}:} This dataset contains $\sim$6.1 hours of multi-party conversations recorded with 360-degree cameras. It faithfully reproduces speech overlap, varying acoustics, and up to 53.7\% silent face segments. We evaluate under three segmentation settings: manual segmentation (Gold), fixed-duration chunks (Fixed Chunk), and active speaker detection (ASD).
\end{itemize}

We adopt word error rate (WER) as the primary metric. 

\subsection{Implementation details}

We adopt a two-stage training strategy. Stage 1 follows the training protocol of the baseline~\cite{nguyen25b_interspeech}\footnote{Our implementation is adapted from the official repository of \cite{nguyen25b_interspeech}: \url{https://github.com/nguyenvulebinh/AVSRCocktail}.}, finetuning the pretrained AV-HuBERT~\cite{shi2022learning} on the combined training set until convergence to ensure a fair comparison. Stage2 freezes the main encoder/decoder parameters and only trains the proposed UBG-Net modules and the Weight-Decomposed Low-Rank Adaptation (DoRA) modules~\cite{liu2024dora} (an improvement over LoRA~\cite{hu2022lora}) inserted into the decoder linear layers ($r=8, \alpha=16$).

The aleatoric uncertainty modeling modules are implemented as 2-layer MLPs with a hidden dimension of 512 and GELU activation. To ensure training stability, the last layer of the semantic projection $\mathrm{MLP}^m_{\mathrm{proj}}$ is initialized to zero. For the Bayesian linear layers, the weights are initialized using Xavier-uniform~\cite{glorot2010understanding} (gain 0.01), and the biases are initialized to -5.0 to encourage sparsity at the start. We optimize the model on 2 NVIDIA 4090 GPUs for 40000 steps using AdamW~\cite{loshchilov2019decoupled} (weight decay 0.005, gradient clipping 1.0) with a peak learning rate of $1e-4$ (linear warmup for 4000 steps). The effective batch size is 64. Additionally, the loss coefficients $\beta_1$ and $\beta_2$ are linearly warmed up for 4000 steps to their target values of $1e-7$ and $1e-4$, respectively, and kept constant thereafter. During inference, the number of Monte Carlo samples is set to $K=5$.

\section{Experimental Results}

\subsection{Comparison with baseline}

We use algorithms in paper~\cite{nguyen25b_interspeech} as the baseline, which can achieve SOTA performance on AVCocktail and LRS2 \footnote{Due to the stochastic nature of our inference strategy, all reported results are averaged over 10 independent runs to ensure statistical reliability.}. Additionally, we include Whisper (large-v3)~\cite{radford2023robust} and Qwen3-Omni (30B-A3B-Instruct)~\cite{Qwen3-Omni} as reference points to contextualize the task difficulty. As shown in the Table~\ref{tab:lrs2_results_fixed} and Table~\ref{tab:wer_and_ablation_study_1}, these general-purpose models suffer severe performance degradation in the presence of interference, underscoring the need for specialized audio-visual adaptation to this challenge.

\noindent \textbf{Robustness on LRS2.}
Table~\ref{tab:lrs2_results_fixed} demonstrates consistent improvements over the baseline across most SNR levels, except for slight declines at 10 dB and under clean conditions, which may be due to over-enhancement. Notably, at -5 dB, the improvement from UBG-Net is more pronounced. This trend confirms the efficacy of our UBG-Net: it explicitly captures signal quality via aleatoric uncertainty to suppress noisy features, while simultaneously mitigating distribution shifts caused by severe corruption through epistemic modeling.

\noindent \textbf{Robustness on AVCocktail.}
As shown in Table~\ref{tab:wer_and_ablation_study_1}, our method outperforms the baseline across all settings. Most notably, in the challenging fixed-chunk setting, we achieve a 7.9\% relative WER reduction. In this setting, the model must handle unsegmented streams containing non-speech frames. We attribute this significant gain to our MUBF mechanism, which acts as a soft voice activity detector. By explicitly modeling signal aleatoric uncertainty, the network effectively downweights multimodal features in the presence of silence or background noise, thereby preventing it from hallucinating to some extent.

\subsection{Ablation studies}

\noindent \textbf{Analysis of modality uncertainty-aware Bayesian fusion (MUBF).}
First, we investigate the contribution of aleatoric and epistemic uncertainty modeling of the proposed MUBF by individually applying them, with the ablation results on the AVCocktail dataset shown in Table~\ref{tab:wer_and_ablation_study_1}. The results in rows 5-6 show that modeling aleatoric or epistemic uncertainty in isolation can improve the baseline to some extent. It is evident that the proposed MUBF mechanism achieves the best performance. This demonstrates the complementary nature of the two uncertainty types: explicitly modeling aleatoric uncertainty serves as a signal-quality indicator that guides the Bayesian gating network to suppress noisy features. Simultaneously, the Bayesian linear models account for parameter variance under distributional shifts. This mechanism effectively calibrates the feature representation and prevents the model from being misled by corrupted inputs, a capability lacking in independent modeling approaches.

\noindent \textbf{Analysis of distribution uncertainty-aware hierarchical voting (DUHV).}
To validate our robust decoding, we compare the proposed DUHV against two alternative sequence-level selection criteria: best confidence (selecting the hypothesis with the highest inference score) and standard majority voting (frequency-based). As shown in Table~\ref{tab:wer_and_ablation_study_1}, our method yields the lowest WER by addressing the limitations of these baselines. While the best confidence strategy suffers from the calibration error inherent in deep models (where the network is often confident but wrong on out-of-distribution samples), and standard majority voting discards valuable likelihood information during ties, our DUHV approach leverages population consensus to filter out inconsistent predictions. Furthermore, by using inference scores as a fine-grained tie-breaker, this approach preserves the model's discriminative power in ambiguous scenarios.

\begin{figure}[t]
  \centering
  \includegraphics[width=1.0\linewidth]{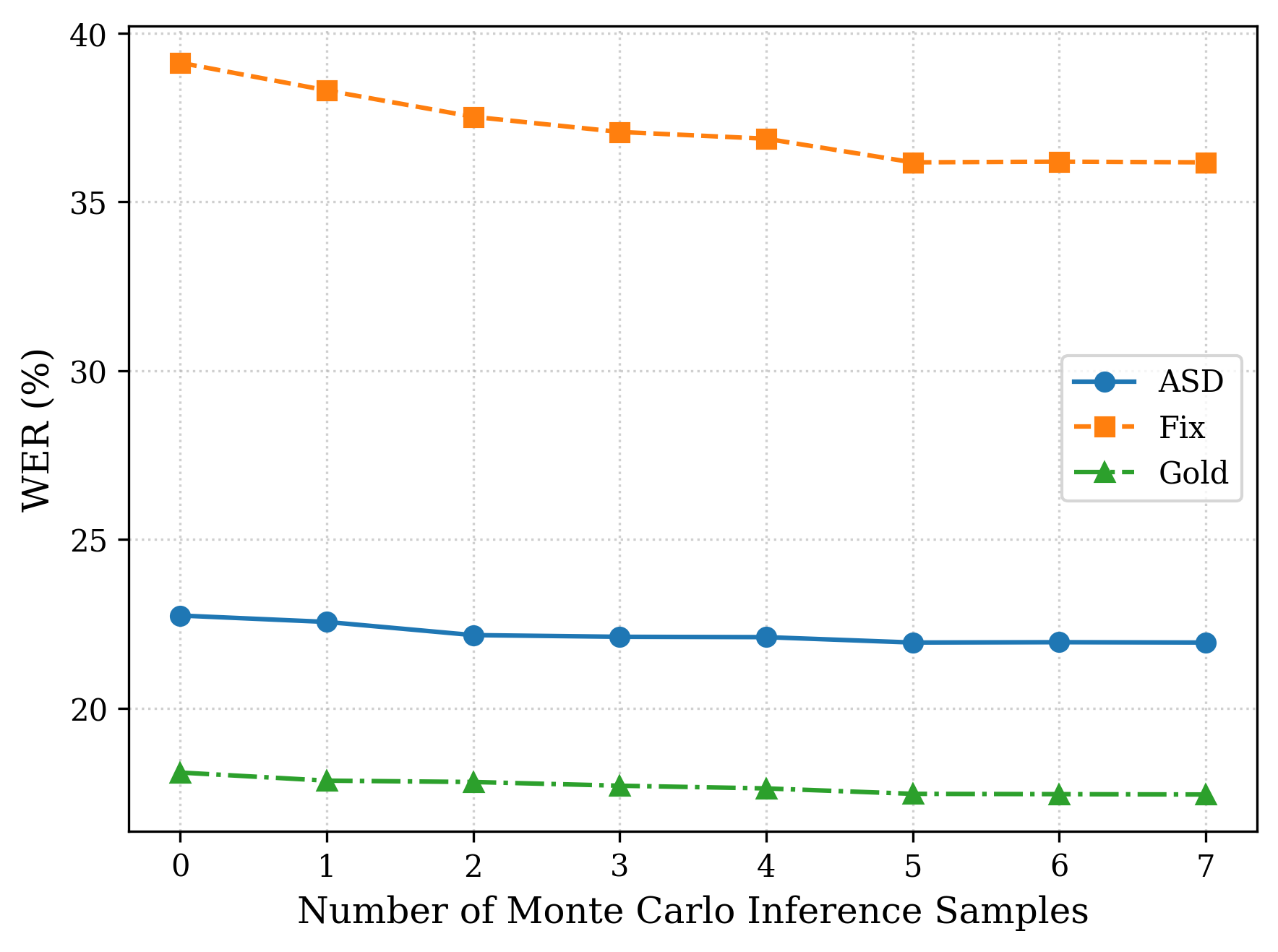}
  \caption{Effect of MC sample size on WER across ASD, Fix, and Gold sets. \textbf{0} denotes deterministic decoding, \textbf{1} represents a single stochastic sample, and $\mathbf{\ge 2}$ indicates voting results.}
  \label{fig:ablation_voting}
\end{figure}

\noindent \textbf{Effect of sample size $K$.}
In the following experiment, we evaluate the parameter $K$ in Monte Carlo (MC) method for the inference. Fig.~\ref{fig:ablation_voting} illustrates the WER trend across different sample sizes, where $K = 0$ denotes standard deterministic beam search (without stochastic sampling). We observe that transitioning from deterministic decoding ($K = 0$) to stochastic sampling ($K \ge 2$) yields significant performance gains, validating the benefit of capturing diverse posterior modes. The WER reduction is most rapid as $K$ increases to 3 and saturates around $K$ = 5. This indicates that $ K=5$ is the optimal operating point for balancing recognition accuracy and inference latency.

\section{Conclusion}

In this work, we proposed a novel uncertainty-aware AVSR framework featuring a Bayesian fusion mechanism that is modality-uncertainty-aware. By explicitly injecting aleatoric uncertainty as a perceptual context to the Bayesian gating network, our method effectively decouples optimization objectives, enabling the epistemic uncertainty modeling module to distinguish irreducible signal noise from semantic ambiguity. Additionally, we introduced a distribution-uncertainty-aware hierarchical voting strategy to robustly resolve sequence alignment errors during Monte Carlo inference. Extensive experiments demonstrate the superiority of our approach, notably achieving a 7.9\% relative WER reduction in the challenging fixed chunk setting of AVCocktail. Future work will explore extending this unified uncertainty paradigm to other multimodal understanding tasks.

\section{Generative AI Use Disclosure}
Generative AI tools were used in a limited manner strictly for language polishing and formatting. No AI tool was used to generate the core scientific contributions, and all authors take full responsibility for the manuscript's content.

\bibliographystyle{IEEEtran}
\bibliography{mybib}

\end{document}